\documentclass[epj,referee]{svjour}
\usepackage{amssymb,amsmath}
\usepackage{latexsym}
\usepackage{graphics}
\usepackage{multirow}
\usepackage{ulem}
\usepackage{color}
\begin{document}
\title{Projectile breakup dynamics for $^{6}$Li~+~$^{59}$Co:
kinematical analysis of $\alpha$-$d$ coincidences}
\author{
F. A. Souza\inst{1}\thanks{\emph{e-mail:} fsouza@dfn.if.usp.br} \and
N. Carlin\inst{1} \and
C. Beck\inst{2} \and
N. Keeley\inst{3} \and
A. Diaz-Torres\inst{4} \and
R. Liguori Neto\inst{1} \and
C. Siqueira-Mello\inst{1} \and
M. M. de Moura\inst{1} \and
M. G. Munhoz\inst{1} \and
R. A. N. Oliveira\inst{1} \and
M. G. Del Santo\inst{1} \and
A. A. P. Suaide\inst{1} \and
E. M. Szanto\inst{1} \and
A. Szanto de Toledo\inst{1}
}                     
%
%
\institute{
Instituto de F\'{i}sica - Universidade de S\~ao Paulo, Departamento de 
F\'{i}sica Nuclear, C.P. 66318, 05315-970, S\~ao Paulo-SP, Brazil \and
Institut Pluridisciplinaire Hubert Curien, UMR 7178, CNRS-IN2P3 et 
Universit\'e Louis Pasteur, Bo\^{i}te Postale 28, F-67037 Strasbourg, 
Cedex 2, France \and Department of Nuclear Reactions, The Andrzej So\l tan 
Institute for Nuclear Studies, ul. Ho\.za 69, 00-681 Warsaw, Poland. \and
Department of Physics, Faculty of Engineering and Physical Sciences,
University of Surrey, Guildford, GU2 7XH, UK
}
\date{Received: date / Revised version: date}
%

\abstract{
A study of the kinematics of the $\alpha$-$d$ coincidences in the 
$^{6}$Li~+~$^{59}$Co system at a bombarding energy of $E_{\rm{lab}} = 29.6$~MeV
is presented. With exclusive measurements performed over different angular
intervals it is possible to identify the respective contributions of the 
sequential and direct projectile breakup components. 
The angular distributions of both breakup components are fairly well described
by the Continuum-Discretized Coupled-Channels framework (CDCC). Furthermore,
a careful analysis of these processes using a semiclassical
approach provides information on both their lifetime 
and their distance of occurrence with respect to the target. Breakup 
to the low-lying (near-threshold) continuum is delayed, and happens at large 
internuclear distances. This suggests that the influence of the projectile 
breakup on the complete fusion process can be related essentially to direct 
breakup to the $^6$Li high-lying continuum spectrum.
} 
\maketitle
%

%
\section{Introduction}
\label{intro}

The breakup process in reactions induced by weakly bound nuclei (such as $^{6}$Li on 
$^{28}$Si~\cite{Pakou06}, 
$^{59}$Co~\cite{Souza09,Souza10}, $^{64}$Zn~\cite{Gomes05}, $^{208}$Pb~\cite{Signorini03}, 
$^{209}$Bi~\cite{Santra09}; 
$^{7}$Li+$^{65}$Cu~\cite{Shrivastava06} and $^{9}$Be+$^{244}$Sm~\cite{Gomes06}) and 
its influence on the fusion cross section 
\cite{Canto06,Keeley07,Canto09,Dasgupta04,Dasgupta99,Tripathi02,Beck03,Diaz02,Diaz03,Marti05,Beck06,Beck07,Beck07b,Beck10}
has been the subject of several experimental and theoretical investigations in recent 
years. In inclusive experiments, 
the light particle spectra measured in `singles' mode display 
significant contributions from reaction mechanisms other than projectile breakup 
\cite{Signorini03,Kelly00,Pakou03}. 
This was also shown very recently for the well studied $^{6}$Li + 
$^{59}$Co system~\cite{Souza09,Souza10}. 
The consideration of either total 
fusion cross sections or complete fusion (CF) cross sections has proved to be
important for a better understanding of the competition between the 
different mechanisms and their respective influence on the fusion process. 
The contributions of sequential projectile  breakup (SBU) and direct 
projectile breakup (DBU) are both significant and it is necessary to determine 
which process influences CF most. In this case, a study of the breakup 
dynamics could provide decisive information. 

The direct breakup DBU seems to be the main cause of the above-barrier CF 
suppression in the $^{9}$Be~+~$^{208}$Pb system, as shown in \cite{Hinde02} 
through sub-barrier measurements of the breakup probability as a function of the 
distance of closest approach. This is a key ingredient for a novel classical
trajectory model with stochastic breakup~\cite{Diaz07} which quantitatively 
relates the breakup process to the ICF and CF cross sections. 

In this work, we present the results of $\alpha$-$d$ coincidence measurements 
(non-capture breakup events) for the $^{6}$Li + $^{59}$Co system at a 
bombarding energy of E$_{\rm{lab}} = 29.6$~MeV, about twice the energy of the 
Coulomb barrier. The same analysis can be applied and similar conclusions may
be drawn for the other 
lower energies studied in our previous work~\cite{Souza09}. However, we have 
chosen to present the results for the highest available energy since it
exhibits the largest measured exclusive cross sections, i.e. the
statistics for binary (projectile breakup and transfer) events are high enough 
to allow a very careful semiclassical analysis. 
By using a simple 3-body kinematics analysis we demonstrate that the 
incomplete fusion (ICF) and/or transfer (TR) processes on the one hand, and 
the projectile breakup components (SBU and DBU) on the other, are associated 
with quite different angular intervals.
In ref.~\cite{Souza09} we presented results from singles measurements which 
show that there is a significant contribution of the ICF/TR processes, i.e.\ 
no distinction between the contributions of ICF and TR was possible. In order 
to gather informations on this important subject, in this work
we also present the results of an investigation of the breakup dynamics, by means 
of calculations related to semiclassical considerations \cite{Tokimoto01} 
involving barrier tunnelling, lifetimes and distances of closest approach for 
the Coulomb trajectories of the projectile and outgoing fragments.
Through this analysis with exclusive measurements we intend to study the 
influence of the breakup on CF by determining the distance of occurrence from 
the target of the SBU and DBU components.

\section{Experimental Setup}
\label{sec:expersetup}

The experiment was performed at the University of S\~ao Paulo Pelletron 
Laboratory, using the 8~UD Tandem accelerator. The $30$~MeV $^{6}$Li beam was 
provided by a SNICS ion source and bombarded a $2.2$~mg/cm$^{2}$ $^{59}$Co 
target. 
It is interesting to note that the energy spectra were not at all affected by 
the relatively large thickness of the target, as shown by Fig.~1, for example.
After correction for energy loss in the target, the effective bombarding 
energy was $E_{\rm{lab}} = 29.6$~MeV (E$_{\rm{cm}}$ = 26.9 MeV) more than twice the energy 
(E$_{\rm{cm}}$ $\approx$ 12 MeV) corresponding to the Coulomb barrier. The beam current 
on target was about $10$~nA. We used 11 triple telescopes \cite{Moura01} for the 
detection and identification of the light particles, positioned on both sides of
the beam axis with 10$^{\circ}$ spacing, covering angular 
intervals from $-45^{\circ}$ to $-15^{\circ}$ and $15^{\circ}$ to $75^{\circ}$. 
The telescopes consisted of an ion chamber with a $150~\mu$g/cm$^{2}$ aluminized 
polypropylene entrance window, a $150~\mu$m thick Si surface barrier detector and 
a CsI detector with PIN diode readout. The ion chamber was operated with $20$~torr 
of isobutane. 

Additional details of the experimental setup and light charged particle 
analysis can be found elsewhere~\cite{Souza09,Moura01}.

\section{Results and discussion}
\label{sec:results}

In a previous publication \cite{Souza09}, we investigated the kinematics of the inclusive $\alpha$ and 
$d$ energy spectra. After subtraction of the estimated 
compound nucleus contributions, broad bumps with significant yields remained in both the $\alpha$ and $d$ 
spectra. The behaviour of the energy centroids of these bumps as a 
function of the detection angle was found to be consistent with dominant 
contributions from incomplete fusion and/or transfer components i.e.:
$\alpha$-incomplete fusion ($\alpha$-ICF)/$\alpha$-transfer ($\alpha$-TR) for the 
$d$ spectra and
$d$-incomplete fusion ($d$-ICF)/$d$-transfer ($d$-TR) for the $\alpha$ spectra. 

These processes are represented, respectively, as follows:\\

$a)$ $^{6}$Li + $^{59}$Co $\rightarrow$ $d$ + $^{63}$Cu$^{*}$
$\rightarrow$ $^{63}$Cu$^{*}$ decay

$b)$ $^{6}$Li + $^{59}$Co $\rightarrow$ $\alpha$ + $^{61}$Ni$^{*}$
$\rightarrow$ $^{61}$Ni$^{*}$ decay\\

The corresponding excitation energies (associated with the energy centroids of the bumps) 
were $24.6$~MeV and $22.5$~MeV for the $^{61}$Ni and $^{63}$Cu nuclei, 
respectively.

\begin{figure}
\centering
\resizebox{0.9\columnwidth}{!}{%
  \includegraphics{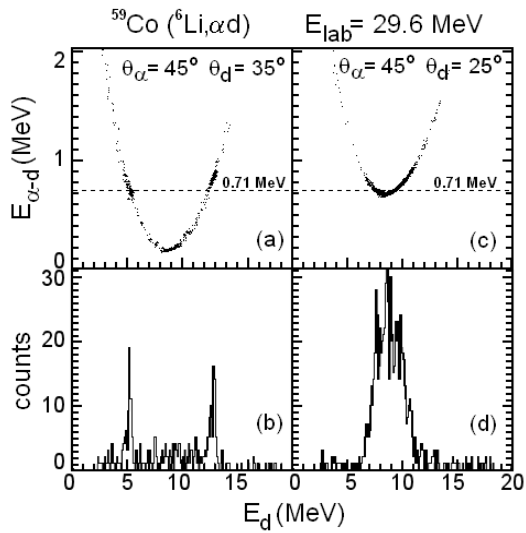}
}
\caption{a) The $\alpha$-$d$ relative energy $E_{\alpha d}$ as a function of the 
deuteron energy for $\theta_{\alpha}=45^{\circ}$ and $\theta_{d}=35^{\circ}$. 
b) The corresponding projection on the deuteron energy axis. c) and d) The same 
for $\theta_{\alpha}=45^{\circ}$ and $\theta_{d}=25^{\circ}$. The dashed lines
correspond to 3-body kinematics calculations assuming $\alpha$-$d$ decays from
the first resonant (2.186 MeV, 3$^{+}$) state of $^{6}$Li.}
\label{fig:CoinSpectra}
\end{figure}

In fig.~\ref{fig:CoinSpectra} typical two-dimensional $\alpha$-$d$ coincidence 
spectra are displayed. 
The large target thickness did not affect the good energy
resolution achieved, shown, for example, by the narrow peaks
in the deuteron spectra of Fig.1.(b).  
For angular differences within the $^{6}$Li 
breakup cone corresponding to the (2.186~MeV, 3$^+$) first resonant state we 
observed two peaks from the two possible kinematical solutions of the SBU. We
also observed a broad structure between the two sharp peaks. 
Alpha-$d$ decay of the second excited state (3.562~MeV, 0$^+$) of $^{6}$Li
is forbidden due to parity considerations and no peak due to the third 
excited state (4.312~MeV, 2$^+$)
was observed. No evidence of decays from higher-lying resonant states was seen. 
For angular differences larger than the SBU cone, we observed only broad 
structures. These broad structures could be associated with either the decay 
of nuclei produced in ICF/TR (incomplete fusion and/or transfer) or 
$^{6}$Li DBU to the continuum. 
For $^{6}$Li~+~$^{198}$Pt, Shrivastava et al.~\cite{Shrivastava09} have 
measured cross sections for $d$ capture (i.e. corresponding to the scenario
previously defined as $b)$ with $d$-ICF) that are much larger than TR cross 
sections.
It is also interesting to note that these non-resonant contributions 
were assumed to arise exclusively from DBU in the case of the 
$^{6}$Li~+~$^{209}$Bi reaction \cite{Santra09} at E$_{\rm{lab}}$ = 36 MeV and 
40 MeV whereas ICF yields were found to represent a large fraction of the 
total reaction cross section in this energy range~\cite{Dasgupta04}. 

\subsection{Kinematics of the $\alpha$-$d$ coincidences}

In order to identify the contributions of the different mechanisms included in 
the broad structures, we performed a 3-body kinematics~\cite{Ohlsen65} analysis of the $\alpha$-$d$ 
coincidence events. We present a study of the $\alpha$ and $d$ energies as a 
function of angle and, as in previous work, we investigate the behaviour of the
energy centroids of the broad structures. For the case of fixed $\alpha$-particle angle, if 
$d$-ICF/TR is dominant the $\alpha$-particle energy should be constant, 
independent of the $d$ emission angle. This energy should be consistent with 
the excitation energy of the intermediate $^{61}$Ni nucleus. Similar behaviour
would be expected for fixed $d$ angle in the case of dominant $\alpha$-ICF/TR; 
the $d$ energy should be constant as a function of the $\alpha$-particle emission 
angle and consistent with the excitation energy of the $^{63}$Cu intermediate 
nucleus. On the other hand, as shown in \cite{Scholz77,Mason92}, if $^{6}$Li 
direct breakup is dominant, the centroid of the broad structure would approximately 
correspond to the minimum allowed $\alpha$-$d$ relative energy for each angular 
pair (see fig.~\ref{fig:CoinSpectra}).

\begin{figure}
\centering
\resizebox{0.9\columnwidth}{!}{
  \includegraphics{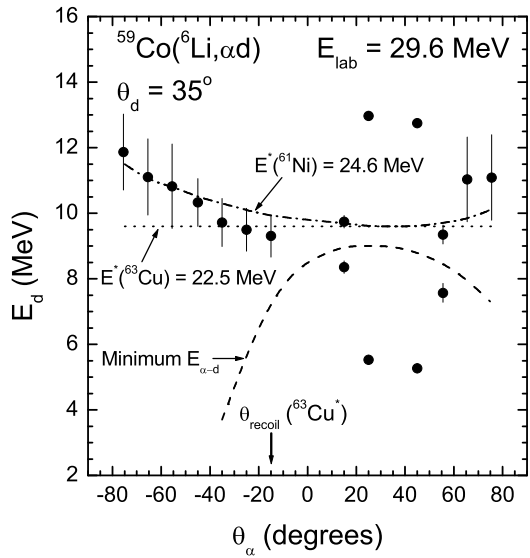}
}
\caption{Experimental values for the deuteron energy as a function of the 
$\alpha$-particle detection angle. The 3-body kinematics predictions for
ICF/TR and the minimum relative energy, $E_{\alpha d}$, for $^{6}$Li breakup 
are also shown.}
\label{fig:Energies}
\end{figure}

In fig.~\ref{fig:Energies} we plot the $d$ energy $E_{d}$ as a function 
of $\theta_{\alpha}$ for $\theta_{d}=35^{\circ}$. In this case, if 
$\alpha$-ICF/TR is dominant, the $d$ energy $E_{d}$ should be constant, 
consistent with the $22.5$~MeV excitation energy of the $^{63}$Cu intermediate 
nucleus (dotted line). This behaviour would be more evident for angles near the 
$^{63}$Cu recoil direction, for which we expect the maximum of cross section for 
the $\alpha$-particle decay. This is indeed observed for angles near the 
recoiling $^{63}$Cu. For other negative angles we observe instead a trend 
consistent with a $24.6$~MeV excitation energy for the $^{61}$Ni composite system
(dot-dashed line). This suggests the dominance of the $d$-ICF/TR process. 
Therefore, both $\alpha$-ICF/TR and $d$-ICF/TR contributions can be, in principle, 
mixed together. The behaviour of the minimum allowed $\alpha$-$d$ 
relative energy for $^{6}$Li breakup is also shown in fig.~\ref{fig:Energies} 
(dashed line). The observed trend suggests that the $^{6}$Li DBU dominates in 
the case of angular pairs for which the broad structure is observed with 
$\Delta\theta_{\alpha d} = 10^{\circ}$ and $20^{\circ}$. For these angular pairs 
(i.e. for $\theta_{\alpha}$ angle values ranging between $10^{\circ}$ and 
$50^{\circ}$) the experimental points shown in fig.~\ref{fig:Energies}
correspond to the energies of the SBU peaks clearly visible in fig.~\ref{fig:CoinSpectra}(b)
and extrapolated in fig.~\ref{fig:CoinSpectra}(d).

\subsection{Breakup dynamics}

In order to gain insight into the dynamics of the SBU and DBU processes, we use a 
semiclassical approximation, following the procedure previously adopted 
in \cite{Kanungo96}. This hypothesis is valid as long as the Sommerfeld parameter 
$\eta$ is large ($\eta\sim6$). High partial waves of the projectile-target
relative motion dominate the non-capture breakup process, so the effect of
the nuclear field on the projectile trajectory is very small. We can then
assume that the projectile travels through the target nuclear field following a Coulomb trajectory. 
This statement is also valid for the breakup fragments, as long as the relative energies are 
not too high.

The relation between the angle of emission and the distance of closest approach is:
\begin{equation}
R_{min}=\frac{Z_{p}Z_{T}e^{2}}{2E}\left[1+\frac{1}{\sin(\theta/2)}\right] \label{eq:Rmin}
\end{equation}
where $Z_{p}$ and $Z_{T}$ are the projectile and target charge numbers, $E$ is the 
centre-of-mass energy and $\theta$ is the scattering angle.

\begin{figure}
\centering
\resizebox{0.9\columnwidth}{!}{
  \includegraphics{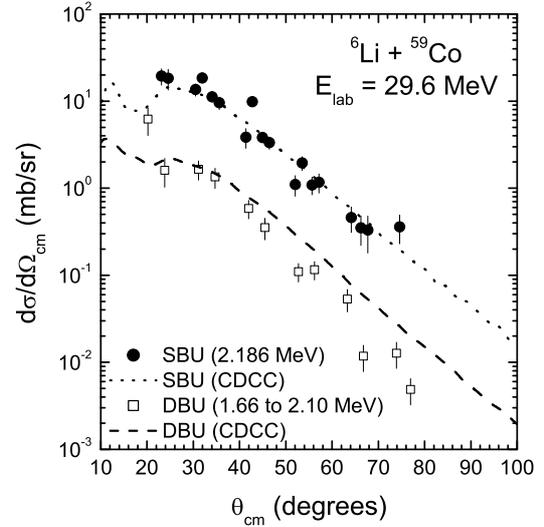}
}
\caption{Experimental angular distributions for the projectile sequential breakup (full circles) 
and direct breakup (open squares) processes, respectively. The dotted 
curve (CDCC calculation \cite{Beck07} for SBU) and dashed curve (CDCC calculation for 
DBU), as described in the text, are used for the semi-classical calculations of
lifetimes and distance of occurrence discussed in fig.~\ref{fig:Functionf} and fig.~\ref{fig:lifetime}.}
\label{fig:AngDistr}
\end{figure}

In order to obtain information on the distances of closest approach related to 
the occurrence of SBU and DBU, we define a quantity $f$ which may
be considered as the relative probability for the production of particles for a given 
process at a given distance of closest approach. The quantity $f$ may be defined as 
follows:
\begin{eqnarray}
f&=&\frac{1}{R_{min}}\frac{{\rm d}\sigma}{{\rm d}R_{min}}\nonumber\\
&=&-\frac{1}{R_{min}}\frac{16\pi E}{Z_{p}Z_{T}e^{2}}\sin(\theta/2)\frac{{\rm d}\sigma}{{\rm d}\Omega} \label{eq:Functionf}
\end{eqnarray}
Here, ${\rm d}\sigma/{\rm d}\Omega$ is the differential cross section in the 
centre-of-mass rest frame for the process under consideration. Figure~\ref{fig:AngDistr} 
depicts the experimental angular distribution for the SBU process 
analyzed in ref.~\cite{Souza09}, as well as for the DBU. The angular distribution 
for the DBU is shown for $^{6}$Li continuum excitation energies summed between 
E$^*$ = $1.66$ MeV and E$^*$ = $2.10$~MeV. In the semiclassical calculations, we 
adopted the most probable value of the excitation energy observed experimentally in this range, 
which is E$^*$ = 1.7 MeV. 
The dotted and dashed lines correspond 
to the SBU and DBU CDCC calculations, similar to those of \cite{Beck07}, respectively.
The $\alpha+d$ binning scheme was appropriately altered to accord exactly with the measured
continuum excitation energy ranges. It should be noted that the calculated elastic scattering 
and the SBU cross sections are unaffected by the change in continuum binning scheme.  
The DBU CDCC result presents a fair agreement with the experimental DBU 
angular distribution at forward angles and a similar shape at backward angles,
although a small difference in the magnitudes is observed.
In table~\ref{tab:xsection} we present the experimental $^{6}$Li SBU cross 
section for the first resonant state (2.186~MeV, 3$^+$)~\cite{Souza09} and 
experimental DBU cross sections for the three excitation energy intervals 
($\Delta E^*$) we observed, together with their corresponding CDCC 
predictions. The CDCC cross sections are in agreement with the experimental 
ones within the uncertainties (that are relatively large for the $E^*$ range 
from 2.20~MeV to 2.40 MeV, as shown in the following) except for the $E^*$ 
range from 3.10~MeV to 3.25~MeV which has a smaller cross section.
In particular, for the DBU cross section corresponding to the $E^*$ range from 
2.20~MeV to 2.40~MeV, we have the situation as shown in 
fig.~\ref{fig:CoinSpectra}(c) and fig.~\ref{fig:CoinSpectra}(d). In this case, 
the SBU is observed in the kinematical limit of the SBU cone, making the 
distinction between SBU and DBU more difficult. In order to overcome this 
problem, we adopted the procedure of calculating the cross section and 
subtracting the corresponding value of the SBU already determined in
previous work~\cite{Souza09} for the situation depicted in 
fig.~\ref{fig:CoinSpectra}(b), where there is no problem with the 
distinction between SBU and DBU. The clear separation of two sharp peaks, as 
well as their widths, in fig.~\ref{fig:CoinSpectra}(b) indicates that the 
broad structure observed in fig.~\ref{fig:CoinSpectra}(d) can not be related 
to effects of energy loss due to the target thickness.
 
\begin{table}
\caption{Experimental SBU and DBU cross sections and CDCC predictions.}
\label{tab:xsection}
\resizebox{1.0\columnwidth}{!}{
\begin{tabular}{c c c c}
\hline\noalign{\smallskip}
    & $\Delta E^*$ (MeV) & $\sigma_{\rm Exp}$ (mb) & $\sigma_{\rm CDCC}$ (mb)  \\
\noalign{\smallskip}\hline\noalign{\smallskip}    
SBU~\cite{Souza09}& 2.186 & $20.6 \pm 4.0$ & 22.1 \\
\noalign{\smallskip}\hline\noalign{\smallskip} 
\multirow{3}{*}{DBU}
    & 1.66 - 2.10 & $3.04 \pm 0.41$ & 2.9 \\
    & 2.20 - 2.40 & $6.8  \pm 4.2$  & 3.5 \\
    & 3.10 - 3.25 & $4.45 \pm 0.94$ & 2.3 \\
\noalign{\smallskip}\hline
\end{tabular}
}
\end{table}

\begin{figure}
\centering
\resizebox{0.9\columnwidth}{!}{
  \includegraphics{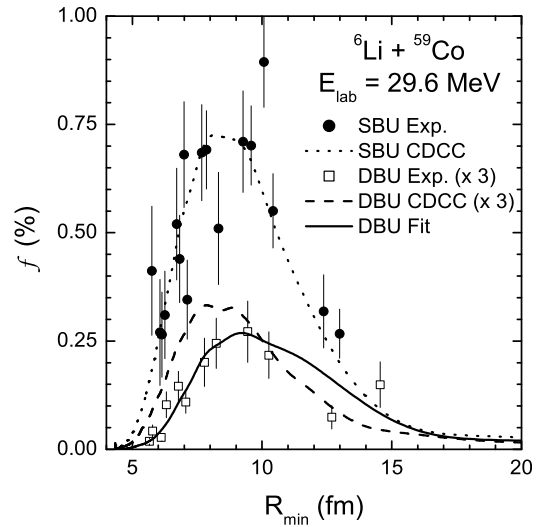}
}
\caption{Function $f$ representing the relative probability for production of particles 
of a given process at a given distance of closest approach, as a function of $R_{\rm{min}}$ for 
the projectile breakup processes SBU and DBU.}
\label{fig:Functionf}
\end{figure}

The curves presented in fig.~\ref{fig:AngDistr} were used for the calculation of 
the $f$ functions shown in fig.~\ref{fig:Functionf} as a function of $R_{\rm{min}}$ for the 
SBU and DBU processes. In particular, a fit to the experimental 
data represented by the solid curve was used for DBU. From the three curves in 
fig.~\ref{fig:Functionf}, the quantity $R_{\rm{min}}^{\rm{MP}}$, the most probable 
value for $R_{\rm{min}}$ for the process of interest, is obtained.

The dip in the angular distributions is probably due to the effect of 
nuclear-Coulomb interference at forward angles ($\sim 20^\circ$). 
This confirms that, as the incident energy is fairly high with respect 
to the Coulomb barrier, the nuclear effects persist to quite forward angles.
In fig.~\ref{fig:Functionf}, this interference effect is associated with
a large $R_{\rm{min}}$ ($\gtrsim 15$~fm). If we took into account the nuclear 
interaction, essentially the tail of the $f$ distribution would be affected. 
Therefore, we expect that the value of $R_{\rm{min}}^{\rm{MP}}$, and consequently the 
conclusions regarding the breakup distances of occurrence will not change.

\begin{figure}
\centering
\resizebox{0.9\columnwidth}{!}{
  \includegraphics{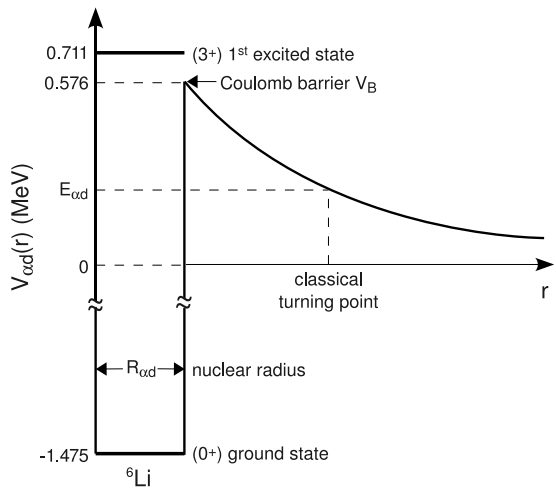}
}
\caption{Schematic representation of nuclear and Coulomb potentials as a function of the
separation distance $r$ between the $\alpha$-particle and the $d$ (adapted from ref.~\cite{Tokimoto01}).}
\label{fig:Potential}
\end{figure}

In this work we also obtained insight into the lifetimes and distance of occurrence from 
the target for the SBU and DBU processes. In particular, for SBU we observed that 
the main contribution is due to the $^{6}$Li $3^{+}$ state with $E^{*} = 2.186$~MeV.
For the DBU, the fragments are no longer bound by the nuclear potential, but 
are still under the influence of the Coulomb barrier between the $\alpha$-particle and 
the $d$. This means that at least for the smaller relative energies, the DBU is a 
delayed process, as is the SBU. A schematic representation of nuclear and Coulomb potentials
is shown in fig.~\ref{fig:Potential} (adapted from ref.~\cite{Tokimoto01}) as a function of the
separation distance $r$ between the $\alpha$-particle and the $d$. The height of the Coulomb 
barrier $V_{\rm{B}}=0.576$~MeV was obtained using $R_{\alpha d}=5.0$~fm and the breakup threshold is defined at $E_{\alpha d}=0$.

In order to estimate the DBU lifetime due to
barrier tunnelling, we adopt the model of \cite{Tokimoto01}, where it was assumed 
that, as in the theory of $\alpha$ decay, the decay rate of the unbound system can be 
written as:
\begin{equation}
 \Lambda_{l}=\omega_{l}P_{l} \label{eq:Decay}
\end{equation}
where $\omega_{l}$ is the barrier bouncing frequency, and $P_{l}$ is the barrier 
transmission probability. Considering that we are dealing with relatively low 
$\alpha$-$d$ relative energies, only the $s$-wave case for which $l = 0$ will be 
considered, for simplicity. In this situation, the bouncing frequency can be estimated 
as being $\omega_{0}=v_{\alpha d}/2R$, where $v_{\alpha d}$ is the $\alpha$-$d$ 
relative velocity and $R$, the nuclear radius. The barrier transmission probability, 
according to the WKB approximation, is given by:
\begin{eqnarray}
P_{0}\approx\sqrt{\frac{V_{\rm{B}}}{E_{\alpha d}}}
{\exp}\bigg\{
-4\eta\bigg[\frac{\pi}{2}
-\arcsin\sqrt{\frac{E_{\alpha d}}{V_{\rm{B}}}}\nonumber\\
-\sqrt{\frac{E_{\alpha d}}{V_{\rm{B}}}\bigg(1-\frac{E_{\alpha d}}{V_{\rm{B}}}\bigg)}~\bigg]\bigg\}
\label{eq:Probability}
\end{eqnarray}
where $\eta=Z_{\alpha}Z_{d}e^{2}/\hbar v_{\alpha d}$ is the Sommerfeld parameter and 
$V_{\rm{B}}=Z_{\alpha}Z_{d}e^{2}/R_{\alpha d}$ is the height of Coulomb Barrier. The lifetime 
can then be determined as $\tau=1/\Lambda$.

Following Coulomb excitation first order perturbation theory \cite{Alder56} and using the time of 
Coulomb excitation as the reference for measuring the lifetime \cite{Tokimoto01}, one can estimate the distance between projectile and 
target when DBU occurs. According to \cite{Alder56,Tokimoto01}, in the focal system 
of the hyperbolic orbit, the distance between projectile and target can be written as:
\begin{equation}
 r=a[\varepsilon\cosh(s)+1]. \label{eq:Distance}
\end{equation}

Here, $a$ is half the distance of closest approach for a head-on collision, $\varepsilon$ is the eccentricity parameter, given by 
$\varepsilon = 1/\sin(\theta^{\rm{MP}}/2)$, with $\theta^{\rm{MP}}$ being the scattering angle 
associated with $R_{\rm{min}}^{\rm{MP}}$. The parameter $s$ \cite{Alder56,Tokimoto01} is related 
to the time $t$ by:
\begin{equation}
t=\frac{a}{v}[\varepsilon\sinh(s)+s], \label{eq:Time}
\end{equation}
where $v$ is the initial relative velocity of projectile and target.

\begin{figure}
\centering
\resizebox{0.9\columnwidth}{!}{
  \includegraphics{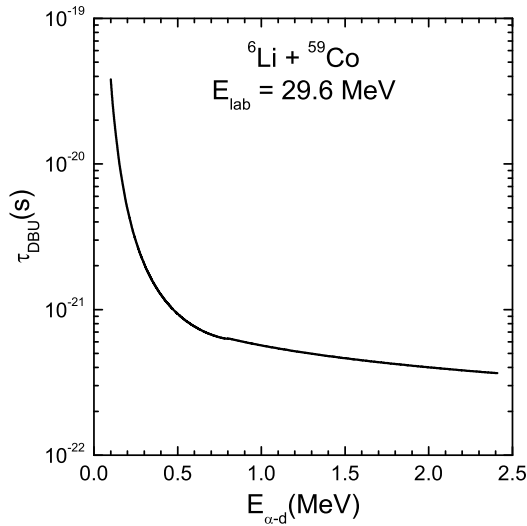}
}
\caption{Calculated lifetime of the DBU continuum states as a function of the relative 
energy $E_{\alpha d}$ from which the corresponding distances of occurrence can be 
deduced in table~\ref{tab:lifetimes}.}
\label{fig:lifetime}
\end{figure}

As described above, from eq.~\ref{eq:Decay} and $l=0$, we can determine the lifetime 
$\tau_{DBU}=1/\Lambda_{DBU}$ for the DBU states near threshold. 
In fig.~\ref{fig:lifetime} we present the plot of $\tau_{DBU}$ ($E_{lab} = 29.6$~MeV) as 
a function of $E_{\alpha d}$. Besides the $^{6}$Li continuum at excitation energy 
$E^{*} = 1.7$~MeV shown in fig.~\ref{fig:AngDistr}, contributions from 
$E^{*} = 2.3$~MeV and $3.2$~MeV were also observed. 
Through eq.~\ref{eq:Time}, and using the $^{6}$Li excitation energies with the corresponding 
values of $\tau_{DBU}$ (in this case $t=\tau_{DBU}$), the values of $s$ can be determined and 
used in eq.~\ref{eq:Distance} to obtain $r_{DBU}$.
The values of $r_{DBU}$ can be compared to those obtained for SBU 
from the $^{6}$Li $3^{+}$ state with $E^{*} = 2.186$~MeV, and knowing that the 
SBU lifetime is $\tau_{SBU} = 2.73 \times 10^{-20}$~s, corresponding to $\Gamma_{SBU} = 
(0.024 \pm 0.002)$~MeV \cite{Tilley02}. The results described above are summarized in 
table~\ref{tab:lifetimes}. Note that the values of $r$ obtained through the two functions $f$ 
for DBU in fig.~\ref{fig:Functionf} are very similar. 

\begin{table}
\caption{Lifetimes, average distance of closest approach and distance of occurrence 
from the target for the projectile breakup components SBU and DBU.}
\label{tab:lifetimes}
\resizebox{1.0\columnwidth}{!}{
\begin{tabular}{c c c c c}
\hline\noalign{\smallskip}    & $E^{*}$ (MeV) & $\tau$ (s)          & $R_{min}^{MP}$ (fm) & $r$ (fm) \\
\noalign{\smallskip}\hline\noalign{\smallskip}
 SBU & 2.18          & $2.7\times10^{-20}$ & $8.5\pm0.4$  & $831.0\pm1.3$ \\
\noalign{\smallskip}\hline\noalign{\smallskip} 
\multirow{3}{*}{DBU}
    & 1.7           & $4.9\times10^{-21}$ & $8.3\pm0.4^{a}$  & $143.1\pm1.5$ \\
    & 1.7           & $4.9\times10^{-21}$ & $9.2\pm0.5^{b}$  & $147.2\pm1.7$ \\
    & 2.3           & $6.3\times10^{-22}$ & $9.1\pm0.7$  & $19.8\pm0.8$ \\
    & 3.2           & $4.4\times10^{-22}$ & $8.1\pm1.2$  & $14.3\pm1.8$ \\
\noalign{\smallskip}\hline
\end{tabular}
}

\begin{footnotesize}
$^{a}$ and $^{b}$: from dashed and solid lines in fig.~\ref{fig:Functionf}, respectively.
\end{footnotesize}
\end{table}

Table~\ref{tab:lifetimes} shows that for all the processes we considered 
the values of $R_{\rm{min}}^{\rm{MP}}$ are very similar. However, the distances of occurrence 
are very distinct for the SBU and DBU. Due to the long lifetime of the resonant 
$^{6}$Li first excited state, sequential projectile breakup occurs very far from the target. 
On the other hand, for DBU the shorter lifetimes of the continuum `states' 
cause the breakup process to occur at shorter distances from the target, although 
there are different distances for different excitation energies in the continuum.

The results obtained in this work are related to the non-capture breakup 
processes as defined in ref.~\cite{Canto06}. However, they can also be 
extended to the case in which only one of the cluster constituents of the 
projectile is captured by the target after projectile breakup. As observed in 
ref.~\cite{Souza09}, from the investigation of the inclusive data, the ICF/TR 
component has been found to have the largest cross section, and therefore has
the major influence on the CF cross section. This conclusion appears to be 
also valid for a heavy target reaction such as $^{6}$Li + $^{209}$Bi 
\cite{Santra09} or $^{6}$Li + $^{198}$Pt \cite{Shrivastava09}. Our results as 
well as refs.\ \cite{Santra09,Shrivastava09} may indicate that the flux 
diverted from CF to ICF would arise essentially from DBU processes via 
high-lying continuum (non-resonant) states of $^6$Li; this is due to the 
fact that both the SBU mechanism and the low-lying DBU processes from low-lying 
resonant $^6$Li states occur at large internuclear distances. The importance 
of ICF (or the so-called fusion suppression 
\cite{Dasgupta04,Dasgupta99,Hinde02}) in reactions induced by weakly-bound 
projectiles is still open. It has been shown recently \cite{Rath09} that for 
a particular projectile ($^6$Li, $^7$Li or $^{9}$Be, for instance), ICF cross 
sections increase with the charge of the target. But, similarly, fusion 
suppression (i.e. ICF) increases with the breakup threshold of the projectile.
Work is still in progress to study ICF processes for $^{6}$Li~+~$^{59}$Co
within the 3-dimensional classical trajectory model of Diaz-Torres and 
collaborators \cite{Diaz07}. 

\section{Summary}
\label{sec:summary}

In a previous study of the reaction mechanisms in the $^{6}$Li~+~$^{59}$Co 
reaction at four different bombarding energies, namely, E$_{\rm{lab}}$ = 17.4 MeV, 
21.5 MeV, 25.5 MeV and 29.6 MeV~\cite{Souza09}, we mainly showed
results of the sequential breakup SBU. In the present work 
we investigated the kinematics for $\alpha$-$d$ coincidences registered 
for the $^{6}$Li + $^{59}$Co reaction at E$_{\rm{lab}}$ = 29.6 MeV, approximately 
twice the energy of the Coulomb barrier. The analysis of the present exclusive 
$\alpha$-$d$ (energy and angular correlations) data along with 3-body 
kinematics calculations allowed us to observe that the ICF/TR 
and the SBU/DBU processes are associated with distinct angular regions. 
The measured breakup (SBU and DBU) cross sections, both consistent with 
Continuum-Discretized Coupled-Channels (CDCC) calculations, are found to
be much larger than for the study with the $^{28}$Si target \cite{Pakou06},
as might be expected from the greater importance of Coulomb breakup for
the $^{59}$Co target. 
The same conclusions can be drawn for the three other lower bombarding
energies. 
In order to complement the results of singles measurements presented in 
ref.~\cite{Souza09}, a semiclassical approach, known to be valid in this  
energy range, was used to estimate the lifetime and distance of 
occurrence with respect to the $^{59}$Co target for SBU and DBU. 
The results indicate that projectile breakup to the low-lying 
(near-threshold) continuum is delayed, and occurs at large internuclear 
distances. To conclude, for $^6$Li + $^{59}$Co the influence of breakup on 
the CF process is essentially due to DBU to the $^6$Li high-lying continuum 
spectrum, i.e., the flux diverted from CF to ICF is due to DBU states.

\begin{acknowledgement}
The authors thank FAPESP and CNPq for financial support.
\end{acknowledgement}

\end{document}